\documentclass{an}
\usepackage{graphicx}
\usepackage{times}
\usepackage{fancyhdr}
\begin{document}
\title{Proper motion Pleiades candidate L-type brown dwarfs}
\author{G. Bihain\inst{1 \and2} \and R. Rebolo\inst{1 \and2} \and V. J. S. B\'ejar\inst{1 \and3} 
\and J. A. Caballero\inst{1} \and C. A. L. Bailer-Jones\inst{4} \and R. Mundt\inst{4}}
\institute{Instituto de Astrof\'isica de Canarias, 38205 La Laguna, Tenerife, Spain 
\and Consejo Superior de Investigaciones Cient\'ificas, Spain
\and Gran Telescopio Canarias Project, Spain
\and Max-Planck-Institut f\"ur Astronomie, K\"onigstuhl 17, 69117 Heidelberg, Germany
}
\date{Received; accepted; published online}

\abstract{We present results of an optical and near-infrared
(IR) 1.8\,deg$^2$ survey in the Pleiades open cluster to search
for substellar objects. From optical $I$-band images from the
CFHT and $J$-band images from the 3.5\,m CAHA Telescope, we
identify 18 faint and very red L~brown dwarf candidates, with
$I>$ 20.9 and $I-J>$ 3.2. The follow-up observations of nine
objects in the $H$- and $K_{\mathrm s}$-bands confirm that eight
belong to the IR sequence of the cluster and the proper motion
measurements of seven candidates confirm that they are Pleiades
members. A preliminary estimation of the substellar mass
spectrum d$N/$d$M$ in the form of a power law $M^{-\alpha}$
provides $\alpha$=$+$0.57$\pm$0.14. We extrapolate this function
to  estimate the number of very low-mass brown dwarfs and
planetary mass objects that could be present in the cluster down
to 1 M$_{\mathrm Jup}$. Sensitive searches combining far red and
near infrared observations may unveal  these objects in a near
future. }

\correspondence{gbihain@ll.iac.es}
\maketitle
\section{Introduction}

After the first discoveries of free-floating brown dwarfs in the
($\sim$120\,Myr, $\sim$130\,pc) Pleiades open cluster (e.g.
Rebolo, Zapatero Osorio \& Mart{\' i}n 1995), fainter objects
were found, with spectral types down to late~M (Zapatero Osorio,
Rebolo \& Mart{\' i}n 1997; Bouvier et al. 1998). Several were
confirmed as Pleiades brown dwarfs by lithium detection (e.g.
Stauffer et al. 1998) or by proper motion (Moraux, Bouvier \&
Stauffer 2001). Then an L0 dwarf, Roque~25 (Mart{\'i}n et al.
1998), and some fainter objects with red near-infrared (IR)
colors were identified, but they still lack a  proper motion or
lithium confirmation (Dobbie et al. 2002; Nagashima et al.
2003). Here we present some Pleiades candidate L-type brown
dwarfs (color $I-J>3.2$) that we confirmed by proper motion (see
Bihain et al. 2005 for more details). Finding L-dwarfs of well
defined ages is necessary to constrain models of stellar and
substellar evolution.

\section{Proper motion candidates}

Using near-IR $J$-band images obtained in 1998 with the 3.5\,m
Telescope/$\Omega$' (Calar Alto Obs.) and optical $RI$-band
images from Bouvier et al. (1998) covering 1.8 deg$^{2}$ of the
Pleiades open cluster, we found 18 L-type low-mass brown dwarf
candidates with magnitudes $I>$ 20.9 and  $J>$ 17.4, and colors
$I-J>$ 3.2. The errors in these magnitudes were $\sigma_{I}\sim$
0.1 and  $\sigma_{J}<$ 0.1, respectively. The data reduction,
the photometry and the coordinates of our candidates are
provided in Bihain et al. (2005). Follow-up $K_{\mathrm s}$~
imaging of eight candidates with the 1.5\,m TCS/CAIN-2 (Teide
Obs.) during winter 2004--2005 allowed us to confirm that seven
belong to the expected near-IR photometric sequence of the
cluster, with colors 1.2 $<J-K_{\mathrm s}<$ 2.0. We obtained
also subarcsecond $H$- and $K_{\mathrm s}$~-band images for
seven objects and the brown dwarf Teide\,1, with the 4.2\,m
WHT/LIRIS (Roque de los Muchachos Obs.) and the 3.5\,m
Telescope/$\Omega$2000 (Calar Alto Obs.) during 2005
January--March. Comparing the pixel positions of the objects in
the first-epoch $I$-band images with those in the $HK_{\mathrm
s}$-band images, we confirmed that they are indeed Pleiades
proper motion members. In the vector point diagram
(Fig.~\ref{lpm}), they all lie at less than 3$\sigma$ (where
$\sigma$ stands for the standard  deviation (8.5 mas/yr) of the
proper motion sample from Moraux et al. 2001) from the cluster
peculiar motion, ($\mu_{\alpha}cos{\delta}$,
$\mu_{\delta}$)=($+$19.15$\pm$0.23, $-$45.72$\pm$0.18)\,mas/yr
(Robichon et al. 1999). We thus confirm that Teide\,1 is a
Pleiades brown dwarf. Observations of this object and the
brightest brown dwarf candidate with the two telescopes allowed
us to obtain double-check measurements  (see linked symbols). 
These are in agreement within $\sim$1\,$\sigma$.

\section{Discussion}

\begin{figure}
\resizebox{0.49\hsize}{!}
{\includegraphics[]{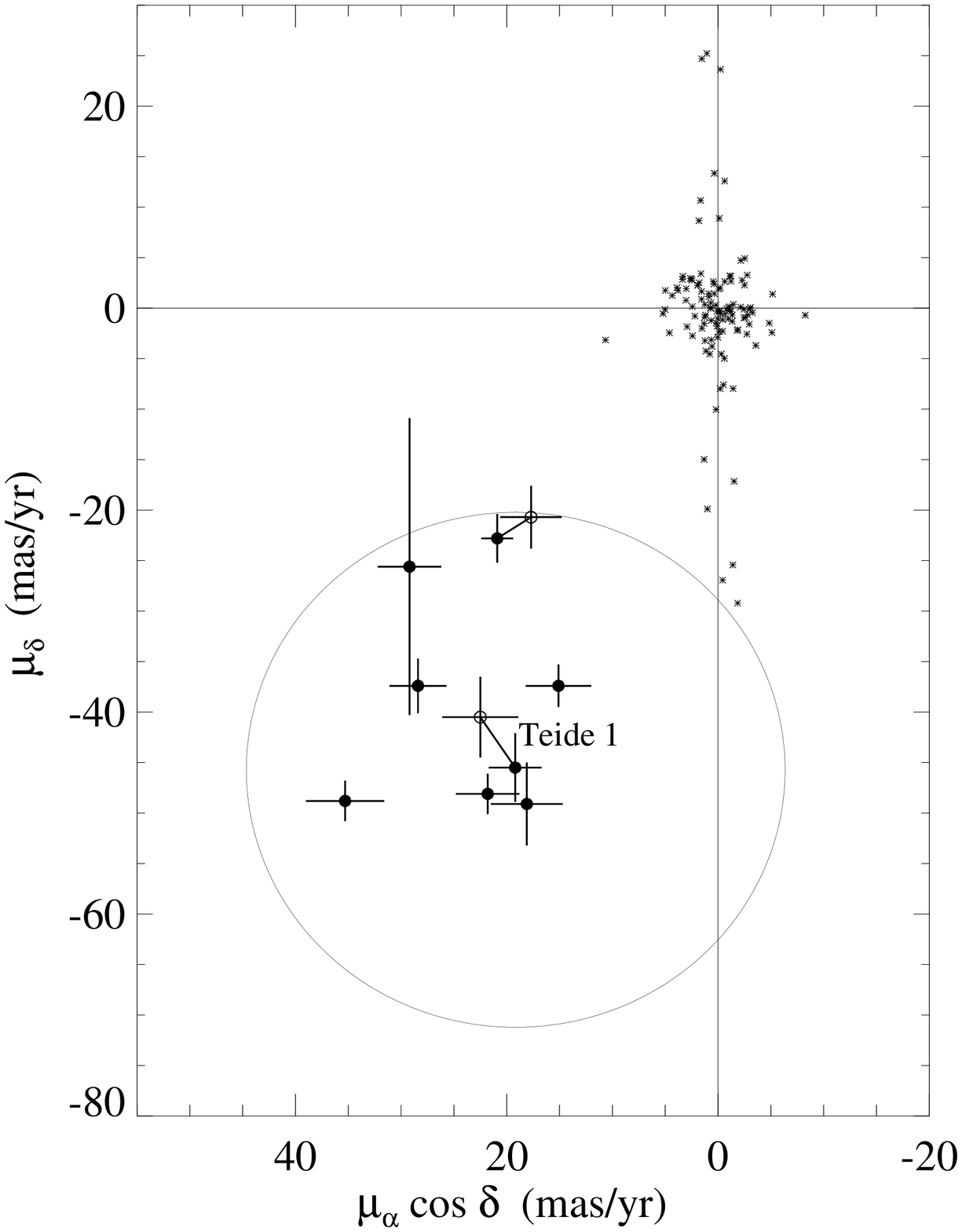}}
\resizebox{0.50\hsize}{!}
{\includegraphics[]{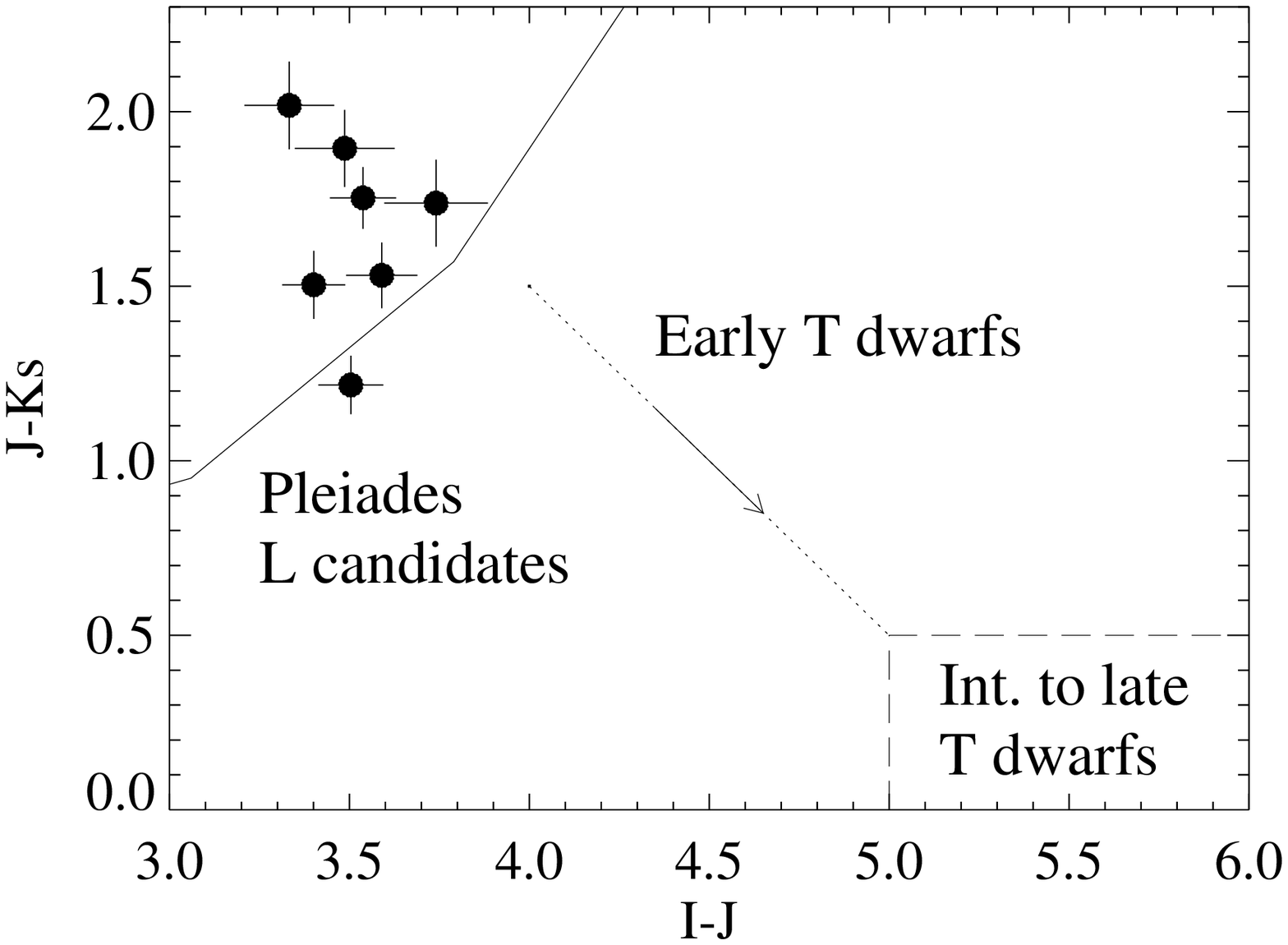}}
\caption{Left: Vector point diagram for the Pleiades candidate L brown dwarfs
(dots with error bars) and the reference objects (asterisks) used
for the astrometric measurements. Right: $J$-$K_{\mathrm s}$, $I$-$J$ diagram for the Pleiades L candidates. The solid line represents the Dusty model isochrone. The regions for the early and the intermediate- to late T dwarfs are indicated.
}
\label{lpm}
\end{figure}

The proper motion candidate L-type brown dwarfs are shown in the
$J-K_{\mathrm s}$,$I-J$ diagram of Fig~\ref{lpm}. The DUSTY
isochrone from Chabrier et al. (2000) is also represented (solid
line), for a cluster age of 120\,Myr and a distance of 133.8\,pc
(Percival, Salaris \& Groenewegen 2005). Most of the L-type
candidates appear bluer in $I-J$ than the model prediction which
takes into account the dust in the brown dwarf atmospheres. This
is possibly due to an underestimation of the $I$-band far red
flux respect to the near-IR flux.

The masses of all our survey candidates were obtained by
comparing their empirical bolometric luminosities (derived from
the sum of their absolute $J$-band magnitude and a bolometric
correction depending of $I-J$) with those from the DUSTY model.
For the L candidates, we obtained masses between 0.040 and
0.020\,M$_{\sun}$. We corrected the number of candidates for the
expected number of foreground and background late-type dwarf
contaminants towards the Pleiades. To compare our results with
previous determinations of the stellar content of the Pleiades,
we  scaled the resulting number  to the whole cluster using an
integrated King profile for a tidal radius 5.54$\degr$
(Pinfield, Jameson \& Hodgkin 1998) and core radii 1.6 and
3.0$\degr$ for stars and brown dwarfs, respectively (Moraux et
al. 2003). Fitting a power law $M^{-\alpha}$ to our mass
spectrum d$N/$d$M$ data points (where d$N$ stands for the number
of objects in the mass range d$M$) and those from Deacon \&
Hambly (2004), and over a mass range from 0.5\,M$_{\sun}$ to the
survey completeness at 0.025\,M$_{\sun}$, we obtained
$\alpha$=$+$0.57$\pm$0.14. 

 A smooth extrapolation of the power law leads to $\sim$100 and
$\sim$70 planetary-mass objects with masses 5--15 and
1--4\,M$_{\mathrm Jup}$, respectively, in the whole cluster. Using
the COND models, the corresponding effective temperatures are
estimated to be in the range  $\sim$600--1300\,K and
$\sim$300--550\,K, respectively.  Objects in the first group are
expected to exhibit T dwarf spectral types, and $I-J$, $J-K_{\rm
s}$ colours as illustrated in Fig~\ref{lpm}. The  $J$-band
magnitudes of Pleiades T dwarfs will be in the range 20--21. 
Most likely, the second  lower mass group will be conformed by
fainter and cooler  objects requiring a new spectral
classification  (Y type?). A combination of deep  I-band imaging
(completeness limit $\sim$26) and near/mid IR observations with
the Spitzer Space Telescope would be an efficient way to  detect
these challenging extremely low-mass Pleiads  in the substellar
realm.

\acknowledgements We are very grateful to J.~Bouvier for
allowing us to use the $RI$-band survey data. We would like also
to thank M.~R.~Zapatero Osorio, A.~T.~Manchado, J.~A.~Pulido,
I.~Villo and J.I.~Gonz{\' a}lez Hern{\' a}ndez for help with the
acquisition of some data.

\end{document}